\documentclass[prb,twocolumn,showpacs]{revtex4}
\usepackage{amsmath}
\usepackage{dcolumn}
\usepackage{graphicx}

\begin{document}

\title[Short title for running header]{Possible origin of the pseudogap end point in the high-T$_{c}$ cuprates}
\author{Tao Li}
\affiliation{Department of Physics, Renmin University of China, Beijing 100872, P.R.China}
\date{\today}

\begin{abstract}
Recent experiments find that the pseudogap phase of the high-T$_{c}$ cuprates ends suddenly at an electron doping $x^{*}$ when the Fermi surface change its shape from hole-like to electron-like. In this short note, we argue that the antiferromagnetic(AF) spin correlation of the system should drop abruptly at the same doping. At the same time, we argue that the critical behavior observed at $x^{*}$ in the specific heat measurement should be attributed to the strong renormalization of the quasiparticle excitation in the anti-nodal region by the critical AF spin fluctuation. This picture also predicts that any pseudogap-like spectral feature related to AF scattering in the electron-doped cuprates should terminate at the doping level when the Fermi surface becomes tangent to the boundary of AF Brillouin zone. Recent ARPES measurement on the electron-doped cuprate Nd$_{2-x}$Ce$_{x}$CuO$_{4}$ seems to be consistent with such a prediction. 
\end{abstract}


\maketitle

Recently, it is found that the pseudogap phase in several hole-doped high-T$_{c}$ cuprates ends suddenly at the electron doping $x^{*}$ when the Fermi surface change its shape from hole-like to electron-like\cite{VHS1,VHS2,VHS3}. More strikingly, it is found that the linear coefficient of the low temperature specific heat exhibits a logarithmic divergence at the same doping\cite{VHS4}. Such a logarithmic divergence is interpreted as the signature of quantum criticality, which is also observed in iron-based superconductor BaFe$_{2}$(As$_{1-x}$P$_{x}$)$_{2}$ and heavy Fermion metal CeCu$_{6-x}$Au$_{x}$ at their antiferromagnetic quantum critical point\cite{QCP1,QCP2}. What is puzzling in the case of the high-T$_{c}$ cuprates is that people do not understand what is really critical at such a high doping level\cite{VHS5}. For example, the antiferromagnetic(AF) order of the system vanishes at a much smaller doping and is thus very unlikely to be the cause of the observed critical behavior. There is also no other obvious ordering channel that is critical at such a doping.

The AF spin fluctuation has been generally believed to be playing an important role in the physics of the high-T$_{c}$ cuprates\cite{NAFL2,NAFL3,NAFL4,NAFL5}. In a series of previous studies, we have shown that the AF band folding effect and the electron pairing caused by the AF spin fluctuation are crucial for the understanding of the quasiparticle dynamics in the pseudogap and the superconducting state\cite{Li1,Li2,Li3}. Unlike the spin fluctuation in weakly correlated metals, which becomes rather weak when the system is tuned away from the magnetic critical point, the spin fluctuation in the high-T$_{c}$ cuprates is robust deep inside the paramagnetic phase as a result of the strong correlation effect.  For example, recent RIXS measurements find that the high energy spin fluctuation in heavily-doped cuprates is hardly changed from that of the AF insulating parent compounds\cite{RIXS1,RIXS2,RIXS3,RIXS4,RIXS5,RIXS6}, indicating its local moment nature.

To account for the dual nature of electron as both local moment and itinerant quasiparticle in the high-T$_{c}$ cuprates, people have introduced the spin-Fermion model, in which the local moment and itinerant quasiparticle behavior are treated phenomenologically as two independent degree of freedoms\cite{NAFL1}.  The spin-Fermion model takes the form of
 \begin{equation}
 H=\sum_{\mathrm{k},\sigma}\epsilon_{\mathrm{k}}c^{\dagger}_{\mathrm{k},\sigma}c_{\mathrm{k},\sigma}+g\sum_{i}\vec{\mathrm{S}}_{i}\cdot\vec{\mathrm{s}_{i}}.
 \end{equation}
Here $\epsilon_{\mathrm{k}}$ is the dispersion of the quasiparticle. $\vec{\mathrm{s}}_{i}=\frac{1}{2}\sum_{\alpha,\beta}c^{\dagger}_{i,\alpha}\vec{\sigma}_{\alpha,\beta}c_{i,\beta}$ is the spin density operator of the itinerant electron. $\vec{\mathrm{S}}_{i}$ is the local moment operator. The dynamics of the local moment is assumed to be described by a phenomenological (inverse) propagator $\chi_{l}(\mathrm{q},\omega)$, which peaks at the AF wave vector Q=$(\pi,\pi)$. $g$ is a phenomenological coupling constant between the quasiparticle system and the local moment system, which is expected to be ferromagnetic\cite{Coupling}.

 \begin{figure}
\includegraphics[width=9cm]{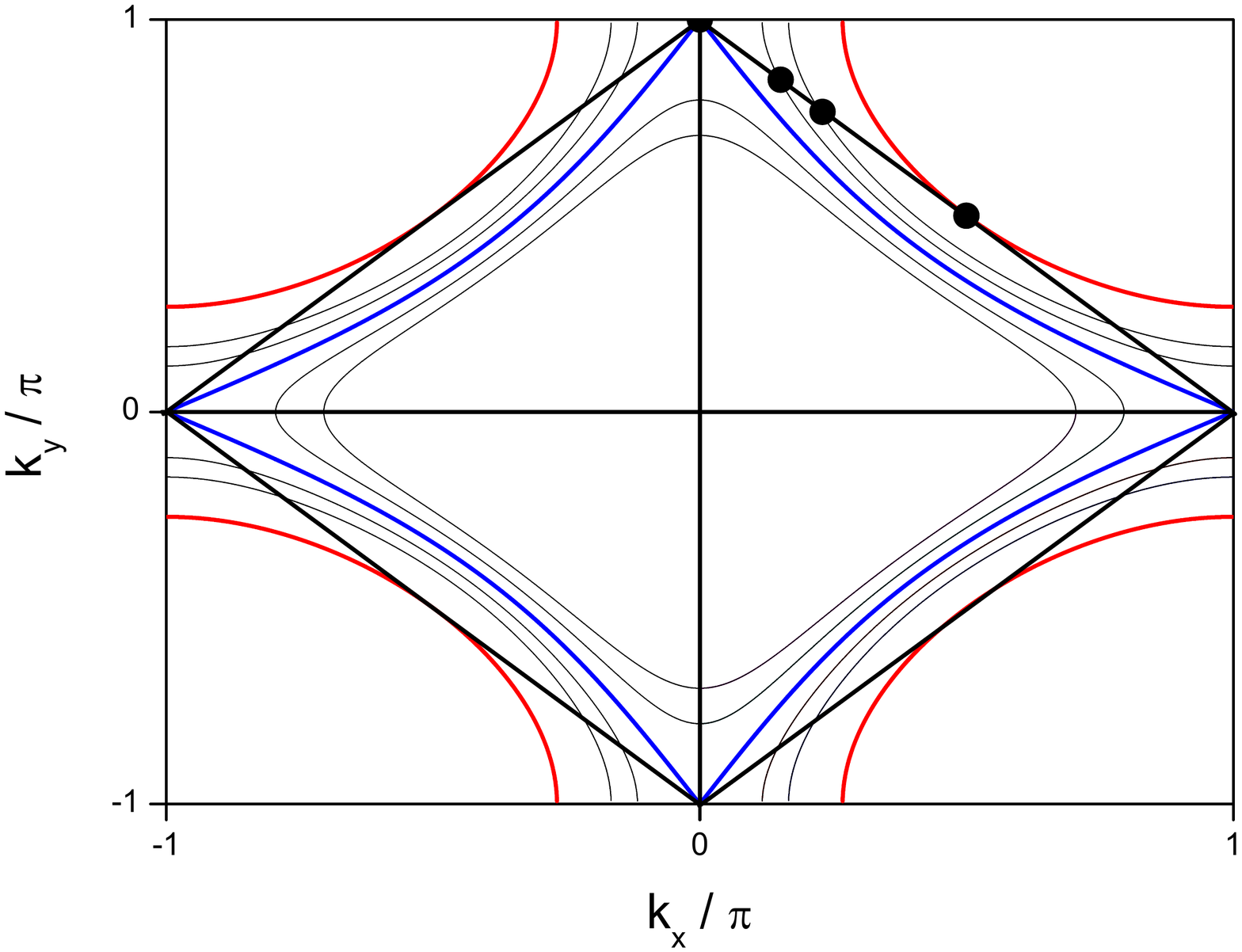}
\caption{\label{fig1}
(Color on-line) The Fermi surface for several different values of $x$ around the pseudogap end point $x^{*}$. The Fermi surface changes its shape from hole-like to electron-like at $x^{*}$. The Fermi surface at $x^{*}$ is denoted by the blue curve. The thick black line marks the boundary of the AF Brillouin zone, which intersects the Fermi surface at the hot spots(denotes by black dots). The hot spots move towards the VHS at M=$(0,\pi)$ or $(\pi,0)$ as we increase $x$ and disappear altogether when $x>x^{*}$. At the electron-doped side of the phase diagram there is another special doping $-x^{0}$, at which the Fermi surface(denoted by the red curve) becomes tangent to the boundary of the AF Brillouin zone. Hot spot exists only for $-x^{0}<x<x^{*}$. Here positive $x$ coresponds to hole doping and negative $x$ corresponds to electron doping.}
\end{figure} 

In the spin-Fermion model, the magnetic susceptibility of the system is determined by the coupled response of the local moment system and the quasiparticle degree of freedom. In the spirit of the random phase approximation, the magnetic susceptibility of the system is given by 
 \begin{equation}
 \chi(\mathrm{q},\omega)=\frac{\chi_{l}+\chi_{i}-2g\chi_{l}\chi_{i}}{1-g^{2}\chi_{l}\chi_{i}}.
 \end{equation}
Here $ \chi_{i}(\mathrm{q},\omega)$ is the bare susceptibility of the itinerant quasiparticles. 
While there is no reason to expect the magnetic response of the local moment system to be sensitive to the change of the Fermi surface shape, the AF response of the quasiparticle system should be singular when the Fermi surface touch the Van Hove singularity(VHS), which are separated from each other by the AF wave vector Q in the momentum space. Here we show that the AF response of the quasiparticle system diverges logarithmically at $x=x^{*}$ and is strongly suppressed when $x>x^{*}$. We argue that the suppression of AF response of the quasiparticle system for $x>x^{*}$ is very likely the origin for the end of the pseudogap phase at $x^{*}$. At the same time, we argue that the critical behavior observed at $x^{*}$ should be attributed to the singular quasiparticle renormalization effect in the anti-nodal region caused by the scattering off critical AF fluctuation. 

To illustrate these points, we calculate the magnetic susceptibility of the quasiparticle system at the AF wave vector as a function of the electron doping. The AF susceptibility of the quasiparticle system is given by
 \begin{equation}
 \chi_{i}(\mathrm{Q},0)=\frac{1}{N}\sum_{\mathrm{k}}\frac{n_{F}(\epsilon_{\mathrm{k+Q}})- n_{F}(\epsilon_{\mathrm{k}})} {\epsilon_{\mathrm{k}}-\epsilon_{\mathrm{k+Q}} },
 \end{equation} 
 in which the quasiparticle dispersion is chosen to be $\epsilon_{\mathrm{k}}=-2t(\cos k_{x}+\cos k_{y})-4t'\cos k_{x}\cos k_{y}-\mu$, with $t'=-0.25t$. $n_{F}(\epsilon)$ is the Fermi function. The Fermi surface for several different doping levels around $x^{*}$ are shown in Figure 1. 

\begin{figure}
\includegraphics[width=9cm]{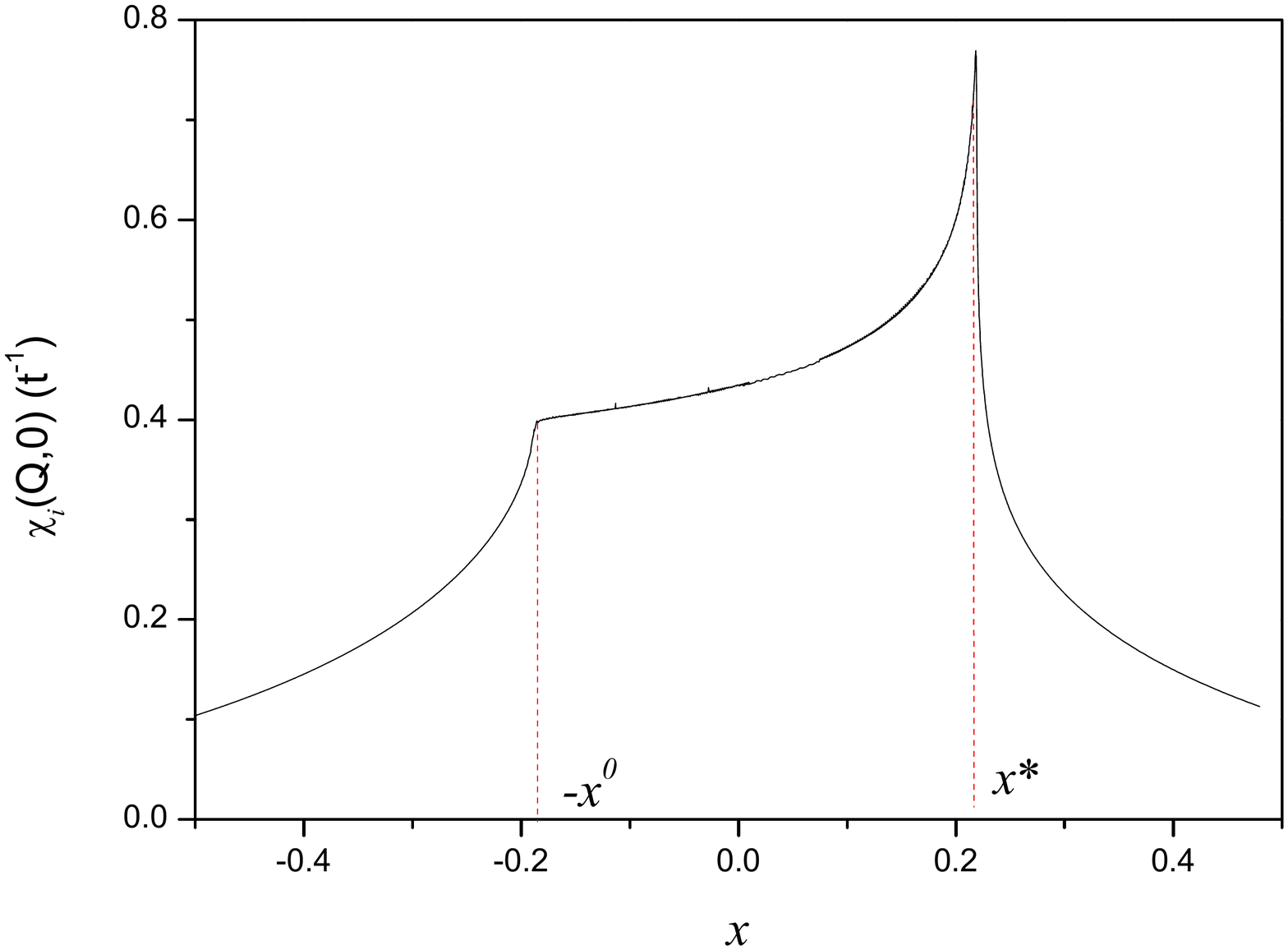}
\caption{\label{fig1}
(Color on-line) The zero temperature magnetic susceptibility of the quasiparticle system at the AF wave vector as a function of electron doping. The red dashed line at $x=x^{*}$ marks the position of the pseudogap end point in the hole-doped cuprates, where the Fermi surface crosses the VHS. The red dashed line at $x=-x^{0}$ marks the position of the pseudogap end point in the electron-doped cuprates, where the Fermi surface becomes tangent to the boundary of the AF Brillouin zone. The quasiparticle system is described by the dispersion $\epsilon_{\mathrm{k}}=-2t(\cos k_{x}+\cos k_{y})-4t'\cos k_{x}\cos k_{y}-\mu$ with $t'=-0.25t$ in this calculation. For such a dispersion relation, we find $x^{*}\approx0.218$ and $x^{0}\approx 0.185$.}
\end{figure}

The hot spots on the Fermi surface, where $\epsilon_{\mathrm{k}}=\epsilon_{\mathrm{k+Q}}$, play an important role in the AF response of the quasiparticle system. From Figure 1 we see that as we increase the hole density the hot spots move towards the VHS and disappear altogether when the Fermi surface crosses the VHS. We thus expect the AF response of the quasiparticle system to be strongly suppressed when the Fermi surface change its shape from hole-like to electron-like. Indeed, the calculated susceptibility at the AF wave vector, which is plotted in Figure 2, drops abruptly for $x>x^{*}$. The logarithmic divergence of the AF susceptibility at $x^{*}$ is caused by the divergence in the effective mass at the VHS.

The local moment system by itself is far away from magnetic criticality for $x=x^{*}$. However, through its coupling to the quasiparticle system, the magnetic response of the local moment system at $x=x^{*}$ will also be driven into critical. At the same time, the abrupt drop in the AF response of the quasiparticle system for $x>x^{*}$ will significantly reduce of AF response of the whole system. If the AF fluctuation is indeed the ultimate origin of the pseudogap phenomena, as we argued for in previous works\cite{Li1,Li2,Li3}, then the suppression of the AF correlation for $x>x^{*}$ is very likely the origin of the pseudogap end point at $x^{*}$. 

At the same time, the divergence of the AF response of the quasiparticle system at $x^{*}$ will greatly enhance its coupling to the local moment system, resulting in divergent self-energy correction for the quasiparticle excitation in the anti-nodal region. We believe this is at the origin of the observed critical behavior in the specific heat coefficient at $x^{*}$. However, we note that the anti-nodal quasiparticle under the scattering of the AF fluctuation is a genuine strongly coupled system as a result of the divergence of effective mass at the VHS. The proximity to the VHS also renders the Migdal theorem strongly violated in the anti-nodal region. These features pose a big challenge for any analytical effort to understand the critical behavior at $x^{*}$.

To make further progress, we now derive a low energy effective theory for the system at $x^{*}$. The action of the effective theory is given by $S=S_{\psi}+S_{\vec{\varphi}}$, in which 
\begin{eqnarray}
S_{\psi}&=&\int d\tau d\mathrm{k}\sum_{\alpha=1,2}\psi^{\dagger}_{\alpha,\mathrm{k}}[\partial_{\tau}-\epsilon_{\alpha}(\mathrm{k})]\psi_{\alpha,\mathrm{k}}\nonumber\\
&+&\frac{g}{2}\int d\tau d\mathrm{k} d\mathrm{q} \ \vec{\varphi}_{-\mathrm{q}}\cdot (\psi^{\dagger}_{2,\mathrm{k+q}}\vec{\sigma}\psi_{1,\mathrm{k}}+\psi^{\dagger}_{1,\mathrm{k+q}}\vec{\sigma}\psi_{2,\mathrm{k}})\nonumber\\
\end{eqnarray}
denotes the effective action of the anti-nodal Fermions. Here we have approximated the Fermion field around the two VHSs at $(0,\pi)$ and $(\pi,0)$ as independent degree of freedoms. $\psi_{1,\mathrm{k}}=(\psi_{1,\mathrm{k},\uparrow},\psi_{1,\mathrm{k},\downarrow})^{T}$ denotes the Fermion field around the $(0,\pi)$ point, whose dispersion is given by $\epsilon_{1}(\mathrm{k})=(t-2t')k^{2}_{x}-(t+2t')k^{2}_{y}$. $\psi_{2,\mathrm{k}}=(\psi_{2,\mathrm{k},\uparrow},\psi_{2,\mathrm{k},\downarrow})^{T}$ denotes the Fermion field around the $(\pi,0)$ point, whose dispersion is given by $\epsilon_{2}(\mathrm{k})=(t+2t')k^{2}_{x}-(t-2t')k^{2}_{y}$. Here the momentum $\mathrm{k}$ is defined with respect to the corresponding VHSs and $\epsilon_{\alpha}(\mathrm{k})$ is approximated to the second order in $\mathrm{k}$. $\vec{\varphi}_{\mathrm{q}}$ is the field of the AF spin fluctuation, whose effective action is given by 
\begin{equation}
S_{\vec{\varphi}}=\int d\tau d\tau' d\mathrm{q} \ \chi^{-1}_{l}(\mathrm{q},\tau-\tau')\vec{\varphi}(\mathrm{q},\tau)\cdot\vec{\varphi}(-\mathrm{q},\tau').
\end{equation}
Here $\chi^{-1}_{l}(\mathrm{q},\tau)$ is the inverse of the dynamical susceptibility of the local moment system. The momentum $\mathrm{q}$ of the spin fluctuation field is now defined with respect to the AF wave vector Q=$(\pi,\pi)$. The effective theory presented above has exactly the same structure as the model proposed by Berg \textit{et al.}\cite{Berg}, which is free of the sign problem in quantum Monte Caro(QMC) simulation. We thus expect that the QMC simulation can also be used to understand the observed critical behavior at $x^{*}$.

While a thorough investigation of the effective action Eq.(4) is beyond the scope of this short phenomenological analysis, a rather sharp prediction can already be made following the logic leading to the criticality of the system at $x^{*}$. More specifically, a close inspection on Fig.1 indicates that there is another special doping at $x=-x^{0}$ in the electron-doped side of the phase diagram, at which the Fermi surface becomes tangent to the boundary of the AF Brillouin zone. For electron doping level $x<-x^{0}$, the hot spots will disappear altogether, just as the case for $x>x^{*}$ in the hole-doped cuprates. We thus expect that if there exist any pseudogap-like phenomena related to AF scattering in the electron-doped cuprates, it should terminate at $x=-x^{0}$. From Fig.2, one find that the magnetic susceptibility of the quasiparticle system at the AF wave vector has a kink singularity at $x=-x^{0}$ and is strongly suppressed for $x<-x^{0}$. Since the singularity of the quasiparticle susceptibility at $x=-x^{0}$ is weaker than that at $x=x^{*}$, we expect a weaker critical behavior at $x=-x^{0}$ in the electron-doped cuprates as compared to that observed at the pseudogap end point of the hole-doped cuprates. 

Indeed, recent ARPES measurement on the electron-doped cuprate Nd$_{2-x}$Ce$_{x}$CuO$_{4}$ find that a pseudogap-like spectral feature disappear at a doping level between $x=0.16$ and $x=0.17$, where the Fermi surface of the system becomes almost tangent to the boundary of the AF Brillouin zone\cite{Shen}. On the other hand, the AF long range order vanishes for $x>0.14$ in this system and the close of the this 'pseudogap' also occurs in the paramagnetic phase.  As compared to the hole-doped cuprates, the doping range of the pseudogap phase in the electron-doped cuprates is much reduced. This may indicate that the pseudogap in both families of cuprates have different origin. As we have argued in Ref.[\onlinecite{Li1}] and Ref.[\onlinecite{Li3}], while the pseudogap in the electron-doped cuprates is very likely an AF band folding gap(an AF band folding gap can survive in the paramagnetic phase if the spin correlation length is large enough and the spin fluctuation is slow enough), electron pairing of some kind must be involved in the formation of the pseudogap in the hole-doped cuprates.

In conclusion, we have argued that the sudden end of the pseudogap phase at $x^{*}$ and the observed critical behavior at the same doping can be understood in the framework of the spin-Fermion model, in which the magnetism is carried by both itinerant quasiparticles and local moments with dominating AF correlation. The same picture also predicts that any pseudogap-like phenomena related to AF scattering in the electron-doped cuprates should terminate at the doping level when the Fermi surface become tangent to the boundary of the AF Brillouin zone. We should also expect that the critical behavior at the pseudogap end point of the electron-doped cuprates to be weaker than that observed in the hole-doped cuprates.


\begin{thebibliography}{99}
\bibitem{Coupling}The ferromagnetic nature of the coupling constant $g$ can be most easily seen from the following identity of the Hubbard interaction, namely, $Un_{i\uparrow}n_{i\downarrow}=-\frac{2U}{3}\vec{\mathrm{s}}_{i}\cdot\vec{\mathrm{s}}_{i}+const$. In the spirit of the auxiliary field path integral formulation, one can interpret one of the spin operator $\vec{\mathrm{s}}_{i}$ as the operator for the local moment.
\end{thebibliography}
\end{document}